\documentclass{svproc}
\usepackage{graphicx}%
\usepackage{multirow}%
\usepackage{amsmath,amssymb,amsfonts}%
\usepackage{mathrsfs}%
\usepackage[title]{appendix}%
\usepackage{xcolor}%
\usepackage{textcomp}%
\usepackage{manyfoot}%
\usepackage{booktabs}%
\usepackage{algorithm}%
\usepackage{algorithmicx}%
\usepackage{algpseudocode}%
\usepackage{listings}%
\usepackage{url}

\usepackage{hyperref}

\newcommand\blfootnote[1]{%
  \begingroup
  \renewcommand\thefootnote{}%
  \footnote{#1}%
  \addtocounter{footnote}{-1}%
  \endgroup
}
\makeatletter
\def\blfootnote{\gdef\@thefnmark{}\@footnotetext}
\makeatother

\begin{document}
\mainmatter              % start of a contribution
\title{Les Misérables Dialogue: A film-based network representation of social communication}
\titlerunning{Les Misérables Dialogue}  % abbreviated title (for running head)

%%% @TODO: remove dagger and orcid when uploading to arxiv
\author{Theodora Moldovan
\and Georgios Panayiotou
\and Davide Vega
}
\tocauthor{Theodora Moldovan, Georgios Panayiotou, and Davide Vega}
\authorrunning{Moldovan, T., Panayiotou, G. and Vega, D.} % abbreviated author list (for running head)

\institute{InfoLab, Dept. of Information Technology, Uppsala University, Uppsala, Sweden\\
\email{\{theodora.moldovan,georgios.panayiotou,davide.vega\}@it.uu.se}\\
$^{\dagger}$ All authors contributed equally to this work.
}

\maketitle              % typeset the title of the contribution

\begin{abstract}
In this work, we revisit the classic Les Mis\'erables benchmark network from the perspective of social communication. 
Using the 1998 film adaptation of the novel, we construct the publicly available \textit{Les Mis\'erables Dialogue} temporal text network, which explicitly distinguishes speakers, addressees, and other recipients of each utterance, along with its content and timing. 
Using the ten social dimensions model, we also derive multilayer network representations of the network, in which layers correspond to distinct relational dimensions. 
Beyond the network dataset, this work offers a general methodology for building similarly rich communication datasets from similar narrative sources.


\end{abstract}

%%%%%%%%%%%%%%%%%%%%%%%%%%%%%%%%%%%%%%%%%%%%%%%%%%%
\section{Introduction}
\label{sec:intro}

%% @TODO: check if this is ok -- ack publishing
\blfootnote{Accepted at COMPLEX NETWORKS 2026. This preprint has not gone any post-submission improvements or corrections.}

% DAVIDE going crazy ---- woop woop
For over a century and half, the story of Jean Valjean - a man who spends decades evading the relentless pursuit of Inspector Javert - has captivated audiences around the world. First told by Victor Hugo in the 1862 novel \textit{Les Mis\'erables}, the story has since been retold on stage, and adapted into numerous films and television series, introducing its cast of characters to generation after generation of readers and viewers. This enduring popularity extends into network science: in 1993, Donald Knuth encoded the co-occurrence of characters across the chapters of the novel as one of the example graphs in \textit{The Stanford GraphBase} \cite{knuth1993stanford}, producing what has since become one of the most widely used toy datasets in the field, appearing routinely in studies of community detection, centrality analysis, and network visualisation.

The popularity of this dataset reflects a broader practice in the study of social systems: relationships are often examined through the structures they produce. In network representations, individuals are represented as nodes and their relationships or interactions as edges, enabling analyses of cohesion, centrality, and community structure. Yet when the phenomenon of interest is social communication, this abstraction captures only part of the process. In the original \textit{Les Mis\'erables} network, for example, an edge indicates that two characters appear in the same chapter, but not whether they speak, who addresses whom, or what is communicated between them. Even communication networks that explicitly connect a sender to a receiver typically collapse an interaction into a single edge, discarding information about \emph{what} was said, \emph{when} it was said, \emph{who} was directly addressed, who else received the message, and how the nature of the interaction changed over time.

Richer representations of communication make it possible to retain this information while preserving the advantages of network modelling. The temporal text network (TTN) model~\cite{vega2018foundations} provides one such representation by explicitly modelling textual messages, the actors who produce and receive them, and the temporal occurrence of each interaction.

In this work, we revisit \textit{Les Mis\'erables} from this perspective of social communication~\cite{goglia2026network}. Using the 1998 film adaptation, we manually reconstruct the dialogue as a TTN in which individual utterances are associated with their speaker, intended addressee, other recipients, textual content, and temporal boundaries. The film medium is particularly well suited to this task because its visual and auditory cues (such as body orientation, gaze, spatial proximity, and tone) provide additional evidence for identifying who participates in a communicative event. More broadly, recent work has also explored feature-rich network representations derived from films~\cite{lafhel2020movie,mourchid2019movienet}.

The textual component of the TTN further allows us to move beyond interaction frequency and examine the \emph{social character} of communication. Individual exchanges can express very different forms of interaction, such as trust, conflict, support, power, or intimacy, even when they involve the same pair of characters. Building on the ten social dimensions proposed by Deri et al.~\cite{deri2018coloring} and the computational model developed by Choi et al.~\cite{choi2020ten}, we associate each utterance with scores describing these dimensions. Using these annotations, we also construct several multilayer social networks~\cite{dickison2016multilayer,kivela2014multilayer}, in which each layer represents a different dimension of social communication.

%  GP: removing below, as this might invite questions about why we focus more on the dataset and not the methodology, and even more so about why movie-based networks are practically useful beyond method benchmarking.
%
%Our goal is to demonstrate a general methodology for constructing richer social communication datasets. 
The main contributions of this work are threefold:
\begin{itemize}
    \item We present a methodology for modelling social communication as a temporal text network that explicitly distinguishes the components of conversational participation ---speakers, addressees, and other recipients--- together with the timing and content of each utterance (Sections~\ref{subse:data} and~\ref{subse:validation}).
 
    \item We construct and make available a richly annotated communication dataset derived from the 1998 film adaptation of \textit{Les Mis\'erables}. In addition to the complete message-level temporal text network, we derive network projections that enable structural, temporal, textual, and multilayer analyses of the same communication system (Sections~\ref{subse:ttn} and~\ref{subsec:mttn})
 
    \item We demonstrate the analytical potential of these representations by comparing their structural properties with the canonical \textit{Les Mis\'erables} network and examining how different forms of social communication and relationships emerge and evolve among the main characters over the course of the film (Section~\ref{sec:analysis}).
\end{itemize}
 
%By reconnecting network structure with the communicative events from which it emerges, this work provides a framework to simultaneously examine who interacts with whom, what kind of social interaction takes place, who participates in it, and how it develops over time.

% TO BE ADDED
% To check: https://academic.oup.com/dsh/article/39/1/74/7429467

% Related work:
% \\ - multilayer network model (layers: character, keyword, location, face, caption) to capture the narration of a movie based on its script, its subtitles, and the movie content + comparative analysis of the 6-movie cycle of the Star Wars saga \cite{mourchid2019movienet}
% \\ - movie similarity through multilayer graph similarity measures \cite{lafhel2020movie}

%%%%%%%%%%%%%%%%%%%%%%%%%%%%%%%%%%%%%%%%%%%%%%%%%%%
\section{Les Mis\'erables: Revisited}
\label{sec:mis}

We construct a communication network from the 1998 film adaptation of \textit{Les Mis'erables}. As in the canonical network, nodes represent characters appearing in the story. The edges, however, encode a fundamentally different relation: rather than linking characters who co-appear within a chapter, they represent directed acts of communication between a speaker and the characters who receive an utterance, either as intended addressees or as secondary listeners.

This construction is methodologically advantageous for a number of reasons. First, the film medium makes every individual present in a scene visible whenever communication takes place, allowing us to encode not only a message's intended addressee, but also secondary recipients who merely witness or overhear it, and how long they remain part of the interaction. Second, it captures the duration and intensity of interactions with a precision that text-only sources cannot offer: a scene's length, or the number of lines exchanged between two characters, can serve as a proxy for the strength of their relationship. Finally, the film's scene structure provides a natural temporal segmentation, enabling dynamic analyses of how relationships change over the course of the film.

At the same time, the collected text of each utterance allows us to infer the social character of the heroes' relationships, and how it evolves over time. As a first case study illustrating the range of analyses this enables, we construct and analyse a multilayer temporal text network whose layers correspond to the ten social dimensions introduced in Section~\ref{sec:intro}.

\subsection{Data collection}
\label{subse:data}

We first constructed a dialogue-level dataset from the the 1998 Les Mis\'erables film\footnote{\url{https://www.imdb.com/title/tt0119683/}}. The dataset contains 1,686 utterances and focuses exclusively on verbal communication between characters. Non-verbal events, such as environmental sounds, music, gestures, or actions not associated with a spoken utterance, are not represented as separate observations. Each utterance constitutes one message and is associated with its textual content, speaker, addressee(s), recipient(s), and start and end times. The temporal boundaries of the messages were obtained from the corresponding subtitle timestamps.

The distinction between \emph{speaker}, \emph{addressee}, and \emph{recipient} is central to the annotation. We operationalize these roles as:
\begin{itemize}
    \item \textbf{Speaker}: the character producing the utterance. Every message has exactly one speaker.

    \item \textbf{Addressee}: the character(s) %, or characters, 
    toward whom the utterance is primarily directed, %. Addressees were 
    identified from % among the characters participating in the scene using
    observable conversational cues such as explicit forms of address,
   % . These included explicit forms of address, 
   the content and % of the utterance and 
   its relationship to the preceding dialogue, the speaker's gaze and body orientation, and the spatial organization of the scene. 
   %In the large majority of cases, a single primary addressee could be identified. A small number of utterances were directed toward more than one character, typically two, while in a few cases no valid addressee could be identified under our operational definition.

    \item \textbf{Recipient}: the character(s) who can reasonably be inferred to have received the utterance, regardless of whether the message was directed specifically toward them. Recipients include the addressee(s)
    and any other character close enough
    %, but can also include other characters participating in the same conversational setting. A character was annotated as a recipient when they were sufficiently close 
    to the speaker to hear the utterance and were observably attending to or participating in the ongoing interaction. 
    %Characters who were present in the scene but engaged in a separate conversation, distracted by another activity, or otherwise not participating in the communicative situation were not considered recipients.
\end{itemize}

This distinction allows the dataset to represent two different aspects of audience structure. The set of addressees captures the \emph{intended direction} of an utterance, whereas the set of recipients captures its \emph{observable communicative reach}. By construction, the set of addressees of a message is always a subset of its recipients. Thus, an utterance directed from character $i$ to character $j$ may simultaneously be received by $j$ and by several other characters who are participating in the same interaction.

In practice, almost all utterances have exactly one addressee. Multiple addressees were assigned only when the communicative target could not be meaningfully reduced to a single character, for example when a speaker addressed a group collectively. A small number of utterances have no addressee. These correspond to cases in which no communicatively available target could be identified according to the criteria above, such as speech occurring in the presence of an incapacitated character. These utterances remain part of the dataset because they are nevertheless spoken messages with an identifiable producer and temporal location.

Annotations relied on the film's complete audiovisual context: in ambiguous cases, annotators drew on tone of voice, gaze, body orientation, and character positioning to determine participation. These cues informed the annotation but were not themselves encoded as messages or edges.

Characters were identified using a consistent naming procedure. When possible, we used the character name given in the film. If no name was provided on screen, we used the name in the closing credits. When neither was available, we used the corresponding name from the novel. Remaining unnamed characters were assigned descriptive labels based on their role or other identifiable attributes (e.g., Soldier 2) while groups of indistinguishable characters, such as groups of soldiers or protesters, were represented using collective descriptive labels.

\subsection{Annotation and validation}
\label{subse:validation}

The initial version of the dataset was produced by a primary annotator~\cite{blomsterberg2025coloring}, who watched the film multiple times while constructing and revising the utterance-level annotations. Because distinctions between addressees, recipients, and characters who are merely present in a scene can occasionally be ambiguous, we subsequently conducted an independent validation procedure. Three validation annotators independently reviewed the resulting annotations. For each utterance, they independently flagged an annotation for revision if they disagreed with the identification of any of its components. 

Table~\ref{tab:agreement} summarises the level of agreement between the primary annotation and the independent validation. For each utterance-component, we report \emph{majority agreement}, the share of cases flagged by no more than one of the three validation annotators, and \emph{unanimous agreement}, the share of cases flagged by none of the three. Overall, the validation shows a remarkably high level of agreement across all three annotation components. Majority agreement exceeds 98\% for every component, and unanimous agreement remains above 90\% throughout.

\begin{table}[t!]
\small
\centering

\begin{tabular*}{\textwidth}{@{\extracolsep{\fill}} lcc}
\toprule
\textbf{Component} & \textbf{Majority} (0-1 flags) & \textbf{Unanimous} (0 flags) \\
\midrule
Speaker & 99.23\% & 97.81\% \\
Addressee(s) & 98.75\% & 94.42\% \\
Recipient(s) & 99.35\% & 90.63\% \\
\bottomrule
\end{tabular*}

\caption{Agreement during the independent validation of the utterance annotations.}
\label{tab:agreement}
\end{table}

Following the independent review, the three validation annotators discussed every utterance that had been flagged by at least one of them. Each flagged case was reconsidered using the corresponding scene and its audiovisual context. The resulting reconciled version, which we refer to hereafter as the \textit{Les Mis\'erables Dialogue dataset}, constitutes the source data used for the network construction described in the following.

\subsection{Temporal text network construction}
\label{subse:ttn}

Starting from the dialogue dataset, we construct the \textit{Les Mis\'erables Dialogue} (\textit{LMD}) temporal text network. A TTN~\cite{vega2018foundations} is represented as a triple $\mathcal{T}=(G, x, t)$, where $G=(A, M, E)$ is a directed bipartite graph consisting of a set of actors $A$, a set of messages $M$, and a set of edges $E$ connecting actors and messages; $x \rightarrow X$ maps each message to its textual content; and $t \rightarrow T$ associates each edge with a temporal annotation. Edge direction represents the flow of information: an edge $(a_i, m_k)$ indicates that actor ($a_i$) produces message ($m_k$), whereas an edge $(m_k, a_j)$ indicates that actor $a_j$ receives it.

We instantiate this existing model directly from the utterance-level annotations described above. The actor set $A$ contains the characters appearing in the dialogue dataset, while the message set $M$ contains one message node for each of the 1,686 utterances.

For every utterance $m_k$ produced by character $a_i$, we create a producer edge $(a_i,m_k)$ whose temporal annotation corresponds to the start time of the utterance. For each character $a_j$ annotated as a recipient, we create a consumer edge $(m_k,a_j)$ whose temporal annotation corresponds to the end time of the utterance. Start and end times are provided by the movie streaming service, which records when each subtitle appears and disappears on screen. Note that these boundaries do not necessarily correspond exactly to the acoustic duration of the underlying speech, but provide a consistent temporal reference throughout the film. We also use the subtitle boundaries as the basis for utterance segmentation, splitting a subtitle into multiple utterances, and therefore multiple message nodes, whenever it contains more than one distinct communicative act. Addressee information is retained separately as a message annotation, allowing us to distinguish the intended target of an utterance from the broader set of characters who receive it.

This construction yields two TTN representations that differ in how message reception is defined. We refer to them as \textit{LMD$_a$} and \textit{LMD$_r$}: in \textit{LMD$_a$}, message-reception edges connect messages only to their intended addressees, whereas in \textit{LMD$_r$} they connect messages to all annotated recipients. Since every addressee is also a recipient, \textit{LMD$_a$} captures direct communication, while \textit{LMD$_r$} additionally represents the broader audience of each utterance.

Finally, we enrich every message in both representations with the ten social dimensions of communication using the classifier from~\cite{choi2020ten}. For each message $m_k\in M$, the classifier produces a vector $\mathbf{s}(m_k)=\left(s_1(m_k),\ldots,s_{10}(m_k)\right)$, where $s_l(m_k)\in[0,1]$ is the score associated with dimension $l$. These scores are retained as message attributes, capturing the social character expressed by each utterance and providing the basis for the multilayer projections introduced below.

\subsection{Derived network representations}
\label{subsec:mttn}

To facilitate the use of the dataset with standard network analysis methods, we additionally derive several character-level networks. These representations differ in (i) whether communication is defined using intended addressees or the broader set of recipients, (ii) whether interaction frequency is retained, and (iii) whether the social dimensions of the messages are considered.

\subsubsection{Flattened projections}
We derive two flattened character-level networks by projecting the message layer $M$ onto the actor layer $A$. Both projections are simple and undirected, and multiple utterances between the same pair of characters are collapsed into a single edge.
In \textit{LMD-flat$_a$}, an edge $(i,j)$ exists if characters $i$ and $j$ are connected by at least one utterance in which one is the speaker and the other is an addressee. %This representation therefore captures direct communication between characters.

The second projection, \textit{LMD-flat$_r$}, is constructed analogously using recipients instead of addressees. An edge $(i,j)$ exists if one character is the speaker of at least one utterance for which the other is annotated as a recipient. Since every addressee is, by construction, also a recipient, \textit{LMD-flat$_r$} includes all edges in \textit{LMD-flat$_a$} together with additional interactions involving other characters who received the utterance. Consequently, $E({\mathrm{\textit{LMD-flat}}_a}) \subseteq E({\mathrm{\textit{LMD-flat}}_r})$.

\subsubsection{Multiplex projections}

Using the message-level social dimension scores, we construct four multiplex projections: two derived from \textit{LMD$_a$} and two from \textit{LMD$_r$}, using thresholds $\theta=0.5$ and $\theta=0.8$. All four networks contain the character set $A$ as actors and ten layers $L$, one corresponding to each social dimension.

For a given threshold $\theta$, an edge $(i,j)$ is created on layer $l$ if there exists at least one message $m_k$ such that (i) $i$ is the speaker and $j$ is connected to the message as an addressee in the \textit{LMD$_a$} projection, or as a recipient in the \textit{LMD$_r$} projection, and (ii) $s_l(m_k)\geq\theta$. An edge on layer $l$ therefore indicates that at least one communication act from $i$ to $j$ expresses dimension $l$ above the selected threshold. Unlike the flattened networks, the multiplex projections retain the original directionality of communication.

%Using the message-level social dimension score $s(m_k)$ we construct two multiplex projections. Both contain the character set $A$ as actors, and ten layers $L$, one corresponding to each social dimension. For a threshold $\theta$, a directed edge $(i,j)$ is created on layer $l$ if there exists at least one message $m_k$ such that both (i) $i$ is the speaker and $j$ is the addressee of $m_k$, and (ii) $s_l(m_k)\geq\theta$. An edge on layer $l$ therefore indicates that at least one communication act between the two characters expresses dimension $l$ above the selected threshold. Unlike the two flattened networks, however, the multiplex projections retain the original directionality of the communication.

Note that a single message may exceed the threshold for multiple dimensions and may consequently contribute to several layers. Conversely, repeated qualifying messages between the same pair of characters are collapsed into a single edge within each layer.

The resulting networks are \textit{LMD-ML$_a^{0.5}$} and \textit{LMD-ML$_r^{0.5}$}, which use the more inclusive threshold $\theta=0.5$, and \textit{LMD-ML$_a^{0.8}$} and \textit{LMD-ML$_r^{0.8}$}, which use the more conservative threshold $\theta=0.8$. Thus, for each threshold we provide both an addressee-based and a recipient-based multiplex representation of the same underlying communication.

%We provide four versions of the multiplex representation, one for each original flattened network. \textit{LMD-ML$_{a}^{0.5}$} and \textit{LMD-ML$_{r}^{0.5}$} use $\theta=0.5$ and provide a more inclusive representation of the social dimensions expressed in the dialogue, whereas \textit{LMD-ML$_{a}^{0.8}$} and \textit{LMD-ML$_{r}^{0.8}$} use the more conservative threshold $\theta=0.8$, retaining only communication acts for which the classifier assigns a higher score to the corresponding dimension. 
%Figure~\ref{fig:distribution_dimensions} shows the original classification scores across the ten dimensions before the thresholding. 

\section{Analysis}
\label{sec:analysis}

To illustrate the range of analyses enabled by the dataset, we first compare the structural properties of the flattened networks with the original \textit{Les Mis\'erables} benchmark, and examine how these properties change when communication is defined using addressees or all recipients. We then explore the character of relationships between the main actors over the course of the film. All analyses are conducted using the \texttt{multinet} package~\cite{magnani_analysis_2021}.

\subsection{Network structure}

\begin{figure}[t!]
    \centering
    \includegraphics[width=.85\textwidth]{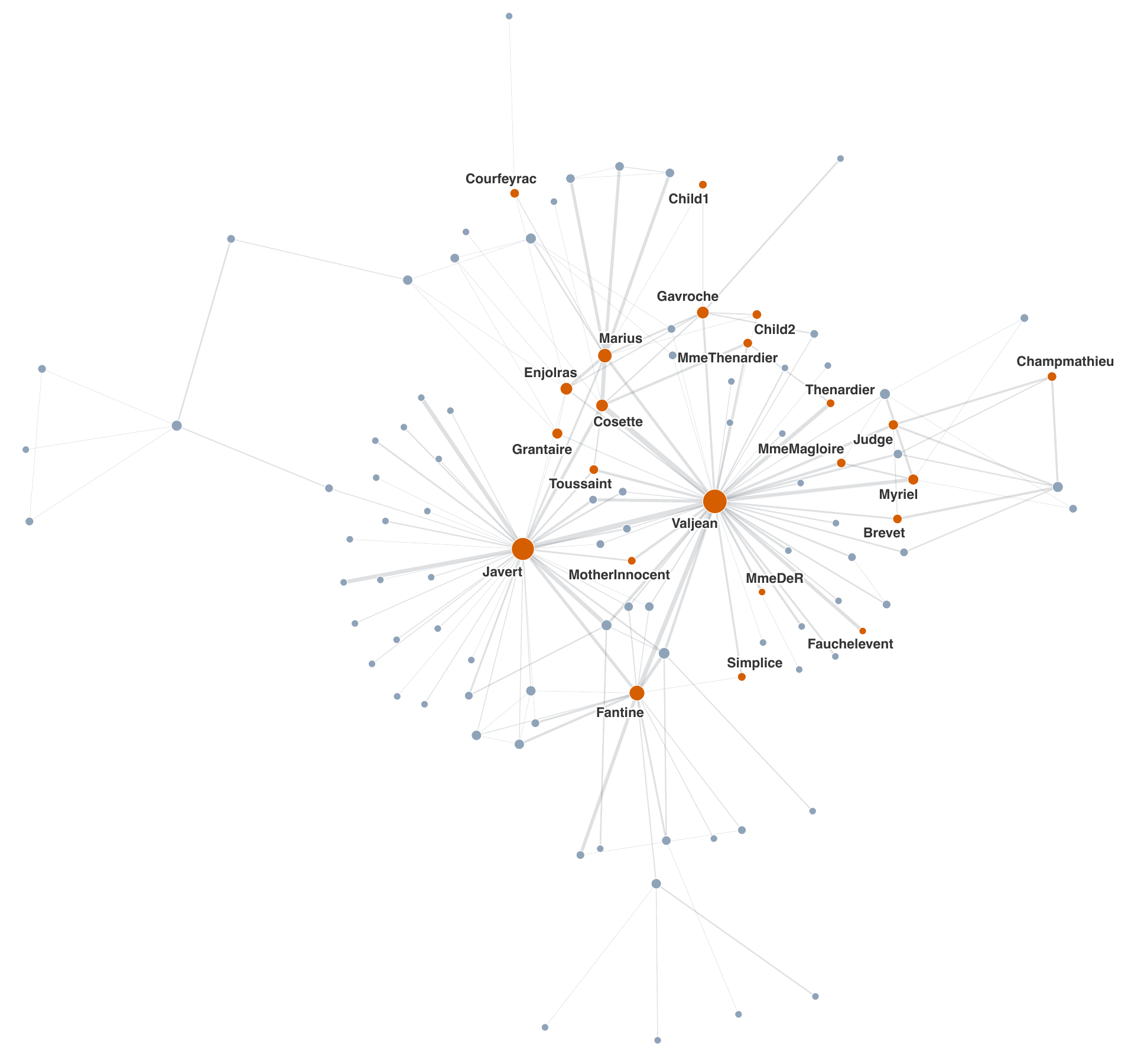}
    \caption{Flattened communication network \textit{LMD-flat}$_a$. The 21 characters also present in the canonical \textit{Les Mis\'erables} network are highlighted and labelled by name.}

    \label{fig:network}
\end{figure}

Table~\ref{tab:net_stats} compares the structural properties of the flattened \textit{LMD-flat} networks with the original \textit{Les Mis\'erables} co-occurrence network. Despite containing more characters, the speaker-addressee network \textit{LMD-flat}$_a$ is sparser, with fewer edges, lower average degree, and substantially lower clustering. Extending the projection to all recipients makes \textit{LMD-flat}$_r$ considerably denser: the number of edges approximately doubles, average degree increases from 3.25 to 5.95, clustering rises from 0.09 to 0.29, and all characters become part of a single connected component.

\begin{table}[t!]
\small
\centering
\begin{tabular*}{\textwidth}{@{\extracolsep{\fill}} l c c l l l l l}

\toprule
\textbf{Network} & \textbf{Nodes} & \textbf{Edges} & \emph{$\langle k \rangle$} & \emph{$\sigma_k$} & \emph{$r$} & \emph{$c$} & \emph{$S$} \\

\toprule
Les Mis\'erables & 77 & 254 & 6.60 & 6.04 & -0.17 & 0.50 & 1.00 \\
\textit{LMD-flat}$_{a}$ & 106 & 172 & 3.25 & 6.18 & -0.40 & 0.09 & 0.91 \\
\textit{LMD-flat}$_{r}$ & 116 & 345 & 5.95 & 8.39 & -0.29 & 0.29 & 1.00 \\
\bottomrule
\end{tabular*}

\caption{Structural properties of the original \textit{Les Mis\'erables} co-occurrence network and the two flattened communication networks derived from our dataset. We report the number of non-isolate nodes and edges, mean degree $\langle k\rangle$, standard deviation of the degree distribution $\sigma_k$, degree assortativity $r$, global clustering coefficient $c$, and relative size of the largest connected component $S$.}
\label{tab:net_stats}
\end{table}

These differences reflect the relations encoded by the networks. The canonical network connects characters that co-occur within chapters, irrespective of direct communication, whereas \textit{LMD-flat}$_a$ requires an explicit communicative act. As we discussed, our dataset includes only speaking characters and is derived from a film adaptation that condenses and reorganises the novel's characters and narrative situations. Accordingly, only 21 characters are shared between the two representations, as highlighted in Figure~\ref{fig:network}. The comparison therefore contrasts two different models derived from different versions and representations of the same artistic work.

The higher connectivity and clustering of \textit{LMD-flat}$_r$ follow naturally from the broader definition of communication that the network encodes. By linking speakers not only to intended addressees but also to other recipients, multi-party scenes introduce additional ties and close triangles that are absent from \textit{LMD-flat}$_a$. This moves \textit{LMD-flat}$_r$ structurally closer to the co-occurrence network, while still requiring characters to participate in the same communicative event rather than merely appear in the same narrative unit.

Both communication networks are disassortative, most strongly for \textit{LMD-flat}$_a$ ($r=-0.40$). This is consistent with highly connected central characters communicating frequently with more peripheral ones. Including all recipients weakens this tendency ($r=-0.29$), as additional ties are introduced among characters participating in the same conversations, although the network remains more disassortative than the canonical co-occurrence network ($r=-0.17$).

\begin{table}[t!]
\footnotesize
\centering
\resizebox{\textwidth}{!}{%
\begin{tabular}{l | l | l | l | l | l | l | l}

\toprule
\textbf{Layer} & \textbf{Edges} & \emph{$\langle k \rangle$} & \emph{$\sigma_k$} & \emph{$r$} & \emph{$c$} & \emph{$S$} & \textbf{d} \\

\midrule
Power      & 12 (\emph{19}) & 0.21 (\emph{0.33}) & 0.74 (\emph{1.12}) & -0.57 (\emph{-0.41}) & 0.00 (\emph{0.05}) & 0.03 (\emph{0.03}) & 2 (\emph{2}) \\
Similarity & 1 (\emph{1})   & 0.02 (\emph{0.02}) & 0.13 (\emph{0.13}) & -- (\emph{--})             & 0.00 (\emph{0.00}) & -- (\emph{--}) & -- (\emph{--}) \\
Conflict   & 27 (\emph{62}) & 0.47 (\emph{1.07}) & 1.50 (\emph{3.05}) & -0.36 (\emph{-0.36}) & 0.10 (\emph{0.08}) & 0.04 (\emph{0.06}) & 3 (\emph{4}) \\
Identity   & 0 (\emph{0})   & 0.00 (\emph{0.00}) & 0.00 (\emph{0.00}) & -- (\emph{--})             & 0.00 (\emph{0.00}) & -- (\emph{--}) & -- (\emph{--}) \\
Support    & 28 (\emph{54}) & 0.48 (\emph{0.93}) & 1.98 (\emph{3.41}) & -0.62 (\emph{-0.53}) & 0.00 (\emph{0.01}) & 0.02 (\emph{0.03}) & 1 (\emph{3}) \\
Knowledge  & 9 (\emph{13})  & 0.16 (\emph{0.22}) & 0.79 (\emph{0.98}) & -0.74 (\emph{-0.49}) & 0.00 (\emph{0.06}) & 0.03 (\emph{0.03}) & 2 (\emph{2}) \\
Fun        & 11 (\emph{25}) & 0.19 (\emph{0.43}) & 0.77 (\emph{1.55}) & -0.06 (\emph{-0.06}) & 0.06 (\emph{0.08}) & 0.02 (\emph{0.03}) & 1 (\emph{2}) \\
Status     & 8 (\emph{16})  & 0.14 (\emph{0.28}) & 0.49 (\emph{0.90}) & 0.00 (\emph{-0.12})  & 0.00 (\emph{0.00}) & -- (\emph{--}) & -- (\emph{--}) \\
Trust      & 45 (\emph{91}) & 0.78 (\emph{1.57}) & 2.33 (\emph{4.02}) & -0.41 (\emph{-0.41}) & 0.07 (\emph{0.08}) & 0.10 (\emph{0.12}) & 4 (\emph{4}) \\
Romance    & 30 (\emph{64}) & 0.52 (\emph{1.10}) & 1.89 (\emph{3.25}) & -0.44 (\emph{-0.38}) & 0.12 (\emph{0.07}) & 0.09 (\emph{0.11}) & 4 (\emph{3}) \\

\bottomrule
\end{tabular}}

\caption{Structural properties of the multiplex network derivations, \textit{LMD-ML$_{a}^{0.8}$} and \textit{LMD-ML$_{r}^{0.8}$} (in parentheses). We report number of nodes and edges, mean degree $\langle k \rangle$, standard deviation of the degree distribution $\sigma_k$, degree assortativity $r$, clustering coefficient $c$, and relative size of the largest strongly connected component $S$.}
\label{tab:mln_stats}
\end{table}

Table~\ref{tab:mln_stats} reports the structural properties of the ten social-dimension layers for the direct addressee network and, in parentheses, the recipient-inclusive network, using the conservative threshold of $0.8$. The layers differ in their prevalence: trust is the most prominent dimension, followed by romance, support, and conflict, while similarity is represented by a single edge and identity by none.

Despite these differences, most layers remain sparse and weakly clustered. Trust and romance form the largest connected components, whereas dimensions such as support and conflict contain a relatively large number of edges distributed across smaller components. The results indicate that some forms of social communication are expressed much more prominently in the dialogue than others, while dimensions with little or no strong signal may be less central to the narrative or conveyed through contextual and non-verbal cues not captured by the spoken text.

\subsection{Temporal dynamics of social dimensions}

\begin{figure}[t!]
    \centering
    \includegraphics[width=\textwidth]{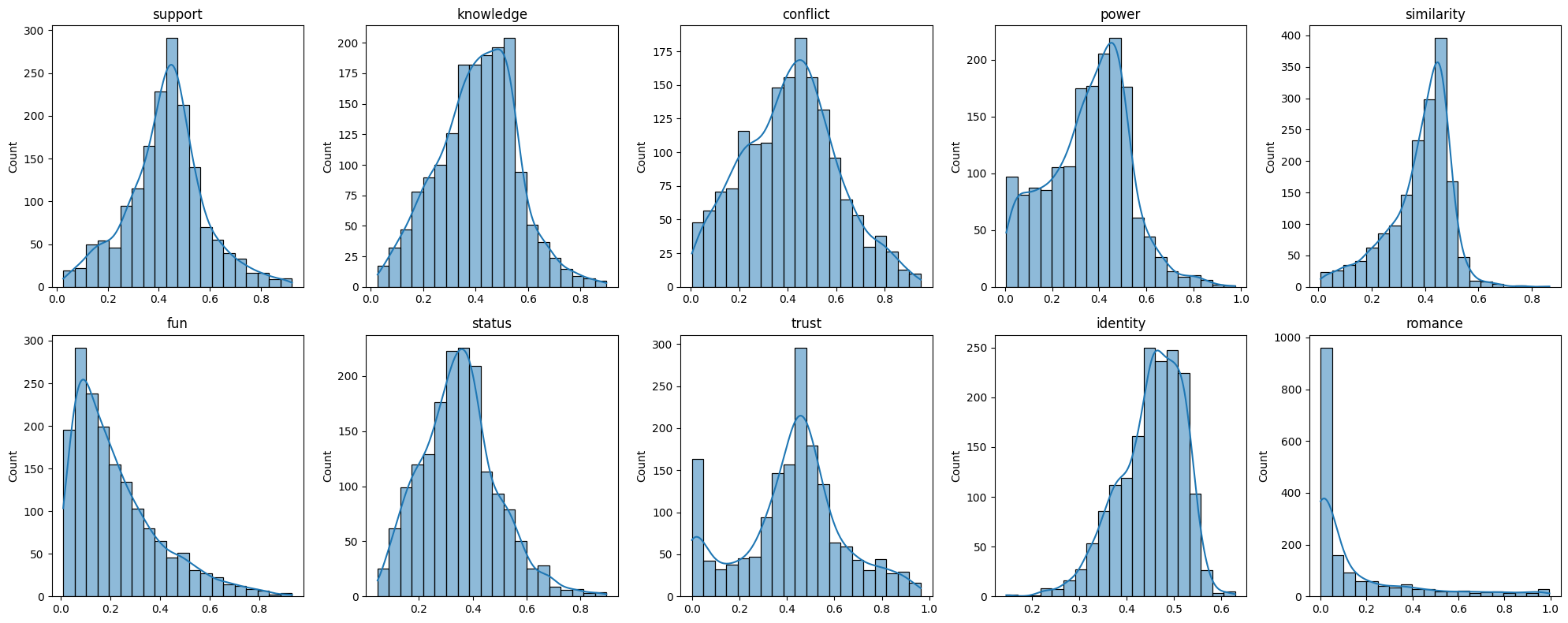}
    \caption{Distribution of classification scores for the ten dimensions across all messages.}
    \label{fig:distribution_dimensions}
\end{figure}

\begin{figure}[t!]
    \centering
    \includegraphics[width=\textwidth]{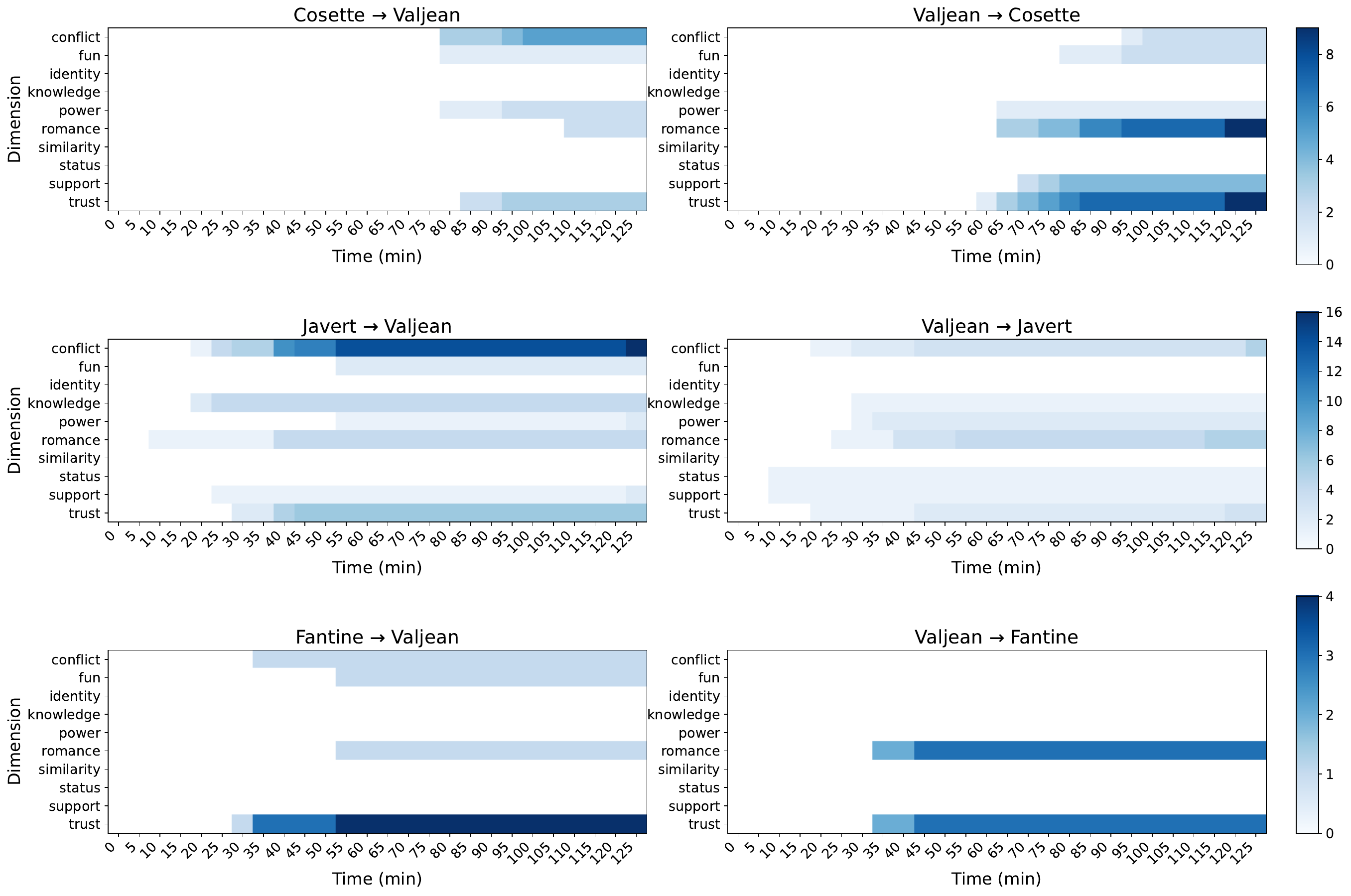}
    \caption{Cumulative message frequency by social dimension over time, for the three most common character pairs. A message counts toward a social dimension if its score for that dimension exceeds 0.8, binned into 5-minute intervals. Color indicates the running total of messages within each bin. Each row shows one dyad, with the two columns giving the two directions of communication (sender $\rightarrow$ receiver).}
    \label{fig:temporal_dimensions}
\end{figure}

To examine how the dominant social dimensions of communication evolve over time, we return to the temporal structure of \textit{LMD$_a$}. For each of the ten social dimensions, we retain only messages between speakers and addressees with a classification score of at least $\theta=0.8$, and preserve their original start times. Figure~\ref{fig:temporal_dimensions} then shows the cumulative frequency of these high-scoring messages over the course of the film for the three character pairs with the largest number of exchanged utterances, with the two directions of communication considered separately.

Cosette and Valjean form the most active pair in the dataset, exchanging 230 messages. In this father-daughter dyad, trust is prominent in both directions, although the profiles diverge over time: Valjean $\rightarrow$ Cosette is dominated by trust and romance, whereas conflict becomes increasingly prominent for Cosette $\rightarrow$ Valjean as she begins to question him. Importantly, a high romance score should not be interpreted as evidence of a romantic relationship between the characters. Instead, it indicates that the language of an utterance contains features associated by the classifier with intimacy or affection, which in this case are consistent with the close familial relationship between Valjean and Cosette.

The 224 messages exchanged between Javert and Valjean are dominated by conflict, particularly in the Javert $\rightarrow$ Valjean direction, consistent with the pursuit that drives much of the film's narrative. Several other dimensions also appear in both directions, but are generally more prominent in Javert's communication toward Valjean.

Fantine and Valjean exchange 89 messages, substantially fewer than the other two pairs. Trust is particularly prominent for Fantine $\rightarrow$ Valjean, consistent with her reliance on him to care for Cosette. Their interaction is also concentrated in the earlier part of the film: the cumulative curves plateau around the first hour, reflecting the comparatively short duration of their relationship following Fantine's death.

%Cosette and Valjean form the most active pair in this dataset, exchanging 230 messages between them. For the father-daughter dyad, trust stands out in both directions, although the profiles diverge over time: Valjean $\rightarrow$ Cosette is dominated by trust and romance, whereas conflict becomes increasingly prominent for Cosette $\rightarrow$ Valjean as she begins to question him. The 224 messages exchanged between Javert and Valjean are dominated by conflict, particularly in the Javert $\rightarrow$ Valjean direction, consistent with the pursuit that drives much of the film's narrative. 

%Several dimensions stand out in both directions of this relationship, but they are more intense for Javert $\rightarrow$ Valjean, reflecting Javert's obsessions with Valjean. For example, trust dominates for Fantine $\rightarrow$ Valjean in the context of asking him to take care of her child. Notably, this relationship's activity peaks early and then flattens, with both directions plateauing by around the first hour mark, consistent with Fantine's death early in the story cutting the relationship short. The 89 messages between Fantine and Valjean are relatively low to the other two pairs discussed above, which further reflects the brevity of the relationship.

%The appearance of fun and romance messages between all three pairs is more puzzling: unlike the other eight dimensions, which have roughly unimodal distributions centred around the middle of the score range, fun and romance scores are heavily skewed toward zero, with only a small fraction of messages scoring high (see Figure~\ref{fig:distribution_dimensions}). This skew makes the 0.8 threshold less reliable for these two dimensions, thus the two dimensions standing out in all relationships may reflect classifier miscalibration rather than genuine signal.

% ML-PSP analysis corroborates results above. 
As a complementary view of these relationships, we finally compute multilayer Pareto shortest paths~\cite{magnani_pareto_2013} on the \textit{LMD-ML$^{0.8}$} networks for the same characters analysed in Figure~\ref{fig:temporal_dimensions}. The results are consistent with the patterns observed above for both the addressee- and recipient-based networks: whenever two characters exchange at least one high-scoring message in a given dimension, they are directly connected on the corresponding layer. When no such direct interaction exists, the shortest connection may instead traverse intermediate characters and multiple social dimensions.

%%%%%%%%%%%%%%%%%%%%%%%%%%%%%%%%%%%%%%%%%%%%%%%%%%%
\section{Conclusion}
\label{sec:conclusion}

In this work, we introduce a methodology for constructing feature-rich communication networks from film dialogue and apply it to \textit{Les Mis\'erables}, producing a richer alternative to the canonical co-occurrence benchmark. The resulting temporal text networks and derived multiplex representations therefore provide not only a new dataset, but also a general framework for studying social communication in narrative media at multiple levels of abstraction.

Several directions remain open for future work. 
First, the temporal resolution of the annotations enables dynamic analyses of how relationships and character roles evolve over the course of the film. 
Second, the recipient structure naturally suggests higher-order representations such as hypergraphs; we provide such a higher-order representation of the dataset in the repository in HIF format~\cite{coll2025hif}. 
The ten-dimensional multiplex representation also invites further comparison of the structural signatures associated with different types of social communication. More broadly, existing methods for analysing such feature-rich networks can be computationally demanding~\cite{panayiotou2024current}, motivating the development of more scalable approaches. We hope that this dataset can serve both as a richer benchmark for network science and as a template for constructing similarly detailed communication networks from other films and narrative sources.

\subsection*{Data availability}
The generated network datasets are publicly available at \url{https://github.com/uuinfolab/les-mis-dialogue}.

% ACKNOWLEDGEMENTS
\subsection*{Acknowledgements}
GP has been partly funded by eSSENCE, an e-Science collaboration funded as a strategic research area of Sweden; and the Anna Maria Lundins Stipendiefond at Smålands nation in Uppsala. The authors also thank Erik Blomsterberg for originally annotating the LMD dataset and the exploratory analysis of the data.

% BIB BIB BIB BIB BIB BIB BIB who's honking at me
\bibliographystyle{spmpsci} % We choose the "plain" reference style
\bibliography{refs} % Entries are in the refs.bib file
\end{document}